# Large bipolarons and oxide superconductivity

David Emin
Department of Physics and Astronomy, University of New Mexico, Albuquerque, NM, USA 87131

## ABSTRACT

Large-bipolaron superconductivity is plausible with carrier densities well below those of conventional metals. Bipolarons form when carriers self-trap in pairs. Coherently moving large-bipolarons require extremely large ratios of static to optical dielectric-constants. The mutual Coulomb repulsion of a planar large-bipolaron's paired carriers drives it to a four-lobed shape. A phonon-mediated attraction among large-bipolarons propels their condensation into a liquid. This liquid's excitations move slowly with a huge effective mass. Excitations' concomitant weak scattering by phonons produces a moderate low-temperature dc resistivity that increases linearly with rising temperature. With falling temperature an energy gap opens between large-bipolarons' excitations and those of their self-trapped electronic carriers.

## 1. INTRODUCTION

Since the late 1940s a form of superconductivity akin to the superfluidity of $^4$He has been conjectured. In analogy with $^4$He atoms, electronic charge carriers in suitable condensed matter would then combine to form bosons, mobile real-space singlet-paired carriers. Pursuing the analogy with liquid helium further, these mobile real-space pairs would then have to condense into a liquid. Finally, such superconductivity requires that with further cooling the liquid undergoes a Bose condensation in order to generate finite occupation of a ground-state which remains fluid.

Evidence of carriers' pairing is offered by the absence of measureable spins in some semiconducting materials with high densities of carriers that move incoherently via low-mobility thermally assisted hopping. However, charge carriers with incoherent transport, small bipolarons, are not expected to provide a suitable basis for superconductivity. Here the requirements for forming coherently moving paired carriers, large bipolarons, are addressed. Large-bipolarons can form in semiconductors with exceptionally displaceable ions as manifested by unusually large ratios of the static to high-frequency dielectric constants, $\varepsilon_0/\varepsilon_\infty \gg 2$.

A large-bipolaron's pair of electronic carriers adjust to phonon-induced alterations of the potential well within which they are self-trapped. This electronic polarization reduces the stiffness constants of the associated atoms. When sufficiently strong, this effect will introduce local vibration modes such as the short-wavelength "ghost modes" which appear in cuprates upon doping. In addition, the cooperative responses of separate pairs of self-trapped electronic carriers lowers the frequencies of phonons whose wavelengths exceed the separation between large bipolarons. This effect generates phonon-mediated attractive interactions between large bipolarons. With a very large static dielectric constant these attractive interactions overwhelm the direct repulsions between large bipolarons thereby enabling their condensation into a liquid.

Large-bipolaron superconductivity would result if the large-bipolarons' ground-state remains liquid rather than condensing further into a solid. By contrast, superconductivity will necessarily be suppressed if the large-bipolaron ground-state becomes globally commensurate with the under-lying square planar lattice of a $CuO_2$ plane. Thus large-bipolaron superconductivity is squelched at doping values of $2/(3\times3)$, $2/(4\times4)$ and $2/(5\times5)$. This feature is consistent with the observed suppression of the superconductivity of $La_2CuO_4$ at a doping of 1/8, in the midst of its superconducting domain.

An atypical frequency-dependent conductivity and low-temperature dc resistivity distinguish a large-bipolaron liquid from conventional conductors. The frequency-dependent conductivity consists of two separate contributions. The slow moving collective excitations of large-bipolarons contribute at applied frequencies below those of the characteristic phonons. By contrast, at applied frequencies above the characteristic phonon frequency, self-trapped carriers are excited from and within their self-trapping potential well. A gap between these low- and high-frequency responses opens as the low-frequency responses shift to lower frequencies with decreasing temperature. The very weak scattering of massive slow moving collective large-bipolaron excitations by long-wavelength acoustic phonons produces a moderate mobility. The concomitant dc resistivity is proportional to temperature even at temperatures well below the Debye temperature.

All told, the large-bipolaron approach to understanding oxide superconductivity only relies on one crucial physical feature, the extremely large ratio of the static to high-frequency dielectric constants of these ferroelectric-like materials. The body of this paper just explains the essential physics underlying this theory. Background material, detailed calculations and discussions of experimental findings are found in longer reviews and in the papers cited below.[1-3]

## 2. SELF-TRAPPING AND POLARON FORMATION

### 2.1 Definitions of self-trapping electronic carrier and a polaron

As illustrated in Fig. 1, for strong enough electron-phonon interactions, it is energetically favorable for electronic charge carriers to *self-trap*. The equilibrium positions of atoms surrounding a carrier are then shifted. The potential well created by these atoms' shifted equilibrium positions binds the carrier with energy $E_{st}$. Energetic stability results when this lowering of the carrier's energy exceeds the energy required to displace the associated atoms. The entity comprising the self-trapped electronic carrier and the associated atomic displacement pattern is termed a (strong-coupling) *polaron*. The frequency with which a self-trapped carrier circulates within its self-trapping potential well then exceeds that of its atoms' vibrations: $E_{st}/\hbar > \omega$.

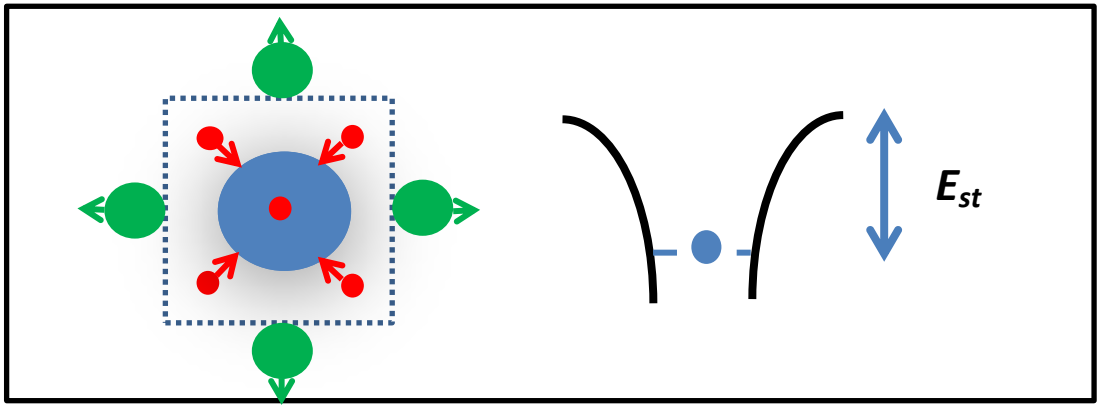

Figure 1. The self-trapping of an electron (blue region) occurs when it is bound with energy $E_{st}$ within the potential well formed by displacing the equilibrium positions of surrounding cations (red dots) and anions (large green dots). Energetic stability requires $E_{st}$ exceed the energy to displace surrounding atoms.

### 2.2 Electron-phonon interactions

An electronic carrier's self-trapping is driven by its electron-phonon interaction. The electron-phonon interaction describes the dependence of the electronic potential experienced by an electronic charge carrier on displacements of a solid's atoms from their carrier-free equilibrium positions. In particular, the carrier's potential energy $V(\mathbf{r})$ is usually taken to depend linearly on atomic displacements:

$$V(\mathbf{r}) = \int d\mathbf{u}\ Z(\mathbf{r} - \mathbf{u})\ \Delta(\mathbf{u}), \tag{1}$$

where $\Delta(\mathbf{u})$ denotes the displacement of the atom whose carrier-free equilibrium position is $\mathbf{u}$. The strength and range of the electron-phonon interaction is then described by the function $Z(\mathbf{r} - \mathbf{u})$. The electron-phonon interaction is generally expressed as the sum of short- and long-range components.

The short-range component of the electron-phonon interaction, similar to a covalent semiconductor's deformation potential, depicts the dependence of the energy of a carrier on displacements of atoms adjacent to it. For a deformable continuum $Z_{SR}(\mathbf{r} - \mathbf{u}) = -F\ \delta(\mathbf{r} - \mathbf{u})$, where $\delta$ denotes the Dirac delta function and $F$ represents the short-range force between the carrier and adjacent atoms. This force is especially strong for oxides because their electronic energies are especially sensitive to oxygen-cation separations. Indeed, as illustrated schematically in Fig. 2, the outer-most electron of an $O^{2-}$ anion is freed upon taking away its surrounding cations.

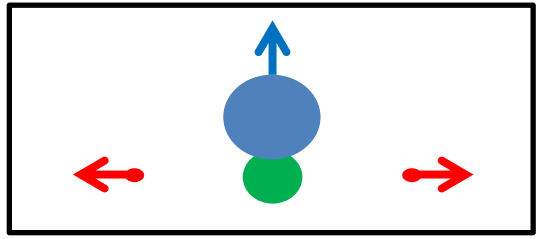

Figure 2. Freeing the outer-most electron (blue region) of an oxygen dianion $O^{2-}$ (large green dot) upon removal of surrounding cations (red dots) generates exceptionally strong electron-phonon interactions.

As depicted in Fig. 3, the long-range component of the electron-phonon interaction results from a carrier's Coulomb interactions with distant ions. This potential can be represented as that from dipoles formed from paired anions and cations. The carrier's potential energy changes as atomic displacements alter the strengths and orientation of these dipoles. The concomitant long-range component of the electron-phonon interaction is:

$$Z_{LR}(\mathbf{r} - \mathbf{u}) = \frac{\sqrt{\frac{e^2}{4\pi}\left(\frac{1}{\varepsilon_\infty} - \frac{1}{\varepsilon_0}\right)\frac{k}{V_c}}}{|\mathbf{r} - \mathbf{u}|^2} cos\theta, \tag{2}$$

where $\theta$ denotes the angle between a dipole located at $\mathbf{u}$ and the vector $\mathbf{r} - \mathbf{u}$, $k$ represents a dipole's Hooke's law stiffness constant and $V_c$ designates the volume of the solid's unit cell. The difference $1/\varepsilon_\infty - 1/\varepsilon_0$ measures the slow polarization generated by movements of ions' cores. Hence, this difference is largest for ferroelectric-like materials containing large concentrations of readily displaceable ions since then $\varepsilon_0 \gg \varepsilon_\infty$.

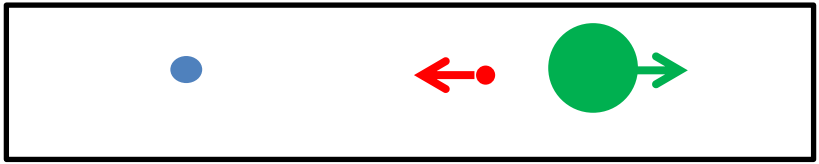

Figure 3. Ionic solids' long-range electron-phonon interactions result from the Coulomb energy of an electronic carrier (blue dot) depending on the positions of distant cations (red dots) and anions (green dots).

**2.2 Self-trapped carrier's non-linear wave equation**

The potential well which binds a self-trapped carrier results from the atoms surrounding it being displaced to new equilibrium positions. Moreover, these displaced atomic equilibrium positions depend on the self-trapped carrier's wave-function. This feedback manifests itself in the self-trapped carrier's ground-state satisfying a non-linear wave-equation.

For simplicity in deriving this non-linear equation, displacements of atoms from their carrier-free equilibrium positions are again presumed to obey Hooke's law while the self-trapped carrier's potential energy is also taken to depend linearly on the displacements of surrounding atoms.

Minimizing the sum of the expectation value of the self-trapped carrier's potential energy and the potential energy of the displaced atoms yields the corresponding pattern of displaced atomic equilibrium positions. These atomic displacements produce the potential well which binds the ground-state polaron's self-trapped carrier. In particular, the non-linear wave-equation for this self-trapped electronic carrier is:[3,4]

$$\left[\frac{-\hbar^2}{2m^*}\nabla_r^2 - \frac{V_c}{k}\int d\boldsymbol{r'}\ |\varphi(\boldsymbol{r'})|^2 \int d\boldsymbol{u}\ Z(\boldsymbol{r}-\boldsymbol{u})Z(\boldsymbol{r'}-\boldsymbol{u})\right]\varphi(\boldsymbol{r}) = E\varphi(\boldsymbol{r}). \tag{3}$$

This equation shows the potential well that binds the self-trapped carrier depending on its wave-function and on the electron-phonon interaction. As a result of this non-linearity, the self-trapped state is qualitatively dependent on the electron-phonon interaction's range and on the electronic dimensionality.[3,4]

**2.3 Scaling analysis of polaron states**

Solving the above non-linear differential equation is generally a very difficult task. However, use of a scaling technique permits the essential features of these solutions to be obtained much more simply.[4] This technique investigates how the ground-state energy of a polaron, the sum of the energy of its self-trapped electronic carrier and the Hooke's law potential energy associated with displacing atoms from their carrier-free equilibrium positions, depends on the self-trapped carrier's spatial extent.

For specificity, attention is now directed to the situation envisioned in cuprate superconductors. That is, the self-trapped charge carriers are envisioned to reside in planes which are surrounded by readily displaceable ions. These electronic charge carriers interact (1) with atoms of their plane via a short-range electron-phonon interaction and (2) with off-plane ions via the long-range electron-phonon interaction.

A strong-coupling polaron's adiabatic energy, that neglecting atoms' vibrations, is the minimum of its energy functional $E_1(R)$ with respect to its (dimensionless) radius $R$. For this planar polaron:

$$E_1(R) = \frac{T}{2R^2} - \frac{U}{2R}\left(\frac{1}{\varepsilon_\infty} - \frac{1}{\varepsilon_0}\right) - \frac{V_{SR}}{2R^2}$$

$$= \frac{(T-V_{SR})}{2R^2} - \frac{U}{2R}\left(\frac{1}{\varepsilon_\infty} - \frac{1}{\varepsilon_0}\right). \tag{4}$$

Here $T$ denotes the electronic carrier's band-width parameter, $V_{SR}$ denotes the energy characterizing the short-range electron-phonon interaction and $U$ denotes the Coulomb interaction between two charges, the energy parameter characterizing the (Coulomb-interaction-based) long-range electron-phonon interaction. The factor of 2 in the denominators of the polaron's potential-energy contributions occurs because with a linear electron-phonon interaction the atomic displacements' strain energy offsets half of the reduction of a carrier's potential energy. The contribution to this energy functional from the short-range component of the electron-phonon contribution varies inversely with the square of $R$ because the charge carriers are taken to be confined to a plane.

Scaling analyses of polaron states shows that they depend critically on the electronic state's dimensionality and on the range of the electron-phonon interactions.[3,4] Indeed, as shown in Fig. 4, there are two distinct types of strong-coupling polaron: small- and large polarons. The scaling expression of Eq. (4) admits both types of solution.

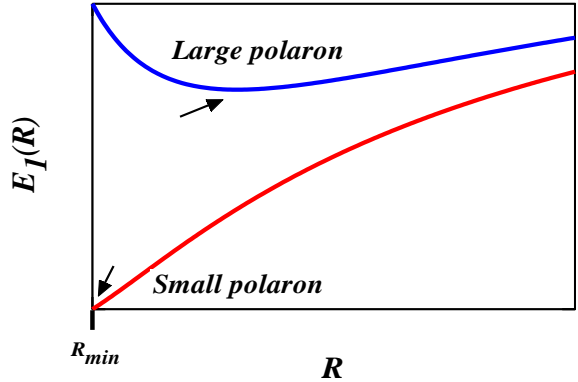

Figure 4. Large-polaron formation (blue curve) has the minimum of its energy $E_1(R)$ as a function of its radius $R$ occurring above its single-site value $R_{min}$. Small-polaron formation (red curve) has the minimum of $E_1(R)$ collapsing to $R = R_{min}$.

A *small* polaron is formed if the minimum of $E_1(R)$ occurs at the smallest allowed value of $R$. Then the self-trapped carrier shrinks to the smallest physically relevant unit (e.g. a single ion, bond or fundamental structural unit). Small polarons move incoherently (by thermally assisted hopping) since the change in energy associated with moving a small polaron between sites generally exceeds the associated transfer energy.[3] For example, with only the short range electron-phonon interaction, small-polaron formation in the planar square lattice described by Eq. (4) requires that $V_{SR} > T$ while semiclassical small-polaron motion loses coherence when $V_{SR}/2 > T/z$, where the number of nearest-neighbors is $z = 4$.

A *large* polaron is generally formed in a multi-dimensional system in the absence of the short-range electron-phonon interaction. Its radius, $a_{Bohr}(m/m^*)/[(1/\varepsilon_\infty) - (1/\varepsilon_0)]$, is typically several times the Bohr radius, $a_{Bohr}$, where $m$ and $m^*$ respectively denote the free-electron and band-effective masses. Large polarons generally move coherently since the transfer energy generally exceeds the change of polaron energy associated with classically moving between sites.[3]

## 3. LARGE-BIPOLARON FORMATION

Large-bipolarons, unlike small-bipolarons, move coherently. Therefore, only large bipolarons can provide a basis for bipolaron superconductivity. The conditions under which large-bipolarons can form are now described.

### 3.1 Continuum model

A bipolaron forms when it is energetically favorable for a pair of electronic carriers to share a common self-trapping potential well. The energy functional governing bipolaron formation differs from that for polaron formation in three respects. First, the sharing of a common potential well by a pair of carriers doubles their net kinetic-energy contribution: $T \to 2T$. Second, the bipolaron contribution to the self-trapping potential energy from two carriers occupying a common state in the self-trapping potential well is quadruple that for polaron formation since each of the two carriers shares a potential well that is doubled: 2 × 2 = 4. Third, the two carriers experience their mutual Coulomb repulsion, $e^2/\varepsilon_\infty |r_1 - r_2|$. As a result the energy functional for a planar bipolaron interacting with off-plane ions becomes:[5,6]

$$E_2(R) = \frac{2T}{2R^2} - 4\left[\frac{U}{2R}\left(\frac{1}{\varepsilon_\infty} - \frac{1}{\varepsilon_0}\right) + \frac{V_{sr}}{2R^2}\right] + \frac{U}{\varepsilon_\infty R}$$

$$= 2E_1(R) + \frac{U}{\varepsilon_0 R} - \frac{V_{sr}}{R^2}. \tag{5}$$

Figure 5 uses plots of $E_2(R)$ and $2E_1(R)$ to explain the circumstances under which an energetically stable large bipolaron forms. By itself, the long-range electron-phonon energy of a bipolaron cancels much of its two carriers' mutual Coulomb repulsion. The inclusion of the short-range component of the electron-phonon interaction can therefore stabilize a bipolaron with respect to forming two separated polarons: $E_2 < 2E_1$. However, the bipolaron will collapse into a small bipolaron if the short-range component of the electron-phonon interaction is too strong. Forming a large bipolaron is therefore limited by the condition[1-3,6]

$$\frac{4\varepsilon_0/\varepsilon_\infty - 6}{(\varepsilon_0/\varepsilon_\infty)^2 - 2} < \frac{2V_{sr}}{T} < 1. \tag{6}$$

The window of acceptable values of the $2V_{sr}/T$ progressively opens as $\varepsilon_0/\varepsilon_\infty$ rises above 2.

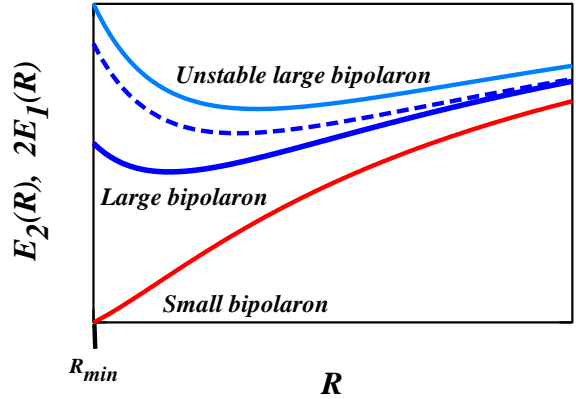

Figure 5. A large bipolaron produced via only an ionic material's long-range electron-phonon interaction is unstable relative to two separate large polarons. In this case the minimum of $E_2(R)$ (uppermost blue curve) exceeds that of $2E_1(R)$ (dashed blue curve). The lower solid blue curve shows the addition of a short-range component of the electron-phonon interaction stabilizing large-bipolaron formation. However, the red curve shows that if the short-range component is too large the bipolaron collapses into a small bipolaron.

Large-bipolaron formation is not even expected for common strongly ionic solids such as KCl where $\varepsilon_0/\varepsilon_\infty \approx 2$. By contrast, the static dielectric constants of the insulating parents of superconducting oxides are very large, $\varepsilon_0/\varepsilon_\infty \gg 2$. These exceptionally large values of the static dielectric constant, characteristic of classic or relaxor ferroelectrics, indicate the presence of readily displaced ions. In particular, the huge static dielectric constants of the cuprates are principally attributed to loosely bound ions that reside between their $CuO_2$ planes. With $\varepsilon_0/\varepsilon_\infty \gg 2$ the large-bipolaron and large-polaron radii approach one another: $R_{bp} \cong R_p \cong a_{Bohr}(m/m^*)\varepsilon_\infty$.

Stabilization of a large-bipolaron is also fostered by reducing the two carriers' mutual Coulomb repulsion. So-called "electron correlation" allows the carrier's two self-trapped electronic states to differ from one another. For a planar large bipolaron with a sufficiently small value of $V_{sr}/T$, these two states are mutually perpendicular dumb-bells.[7] As shown in Fig. 6, the resulting planar large bipolaron assumes a four-lobed shape.

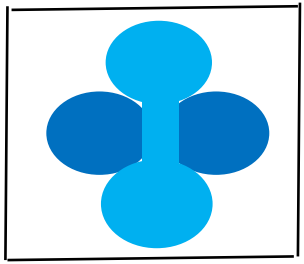

Figure 6. The Coulomb repulsion between a planar large-bipolaron's two electronic carriers is reduced by their assuming the mutually perpendicular dumbbell-shaped states indicated by different shades of blue.

### 3.2 Microscopic model

On the microscopic scale, I conjecture a planar large bipolaron formed from two holes added to a $CuO_2$ plane of $La_2CuO_4$.[3] For example, Fig. 7 illustrates an hypothesized model.[3] I envision the core of this planar large bipolaron to involve four oxygen anions bounded by a square of copper cations. The hole-type bipolaron removes two of the eight out-of-plane electrons from the four bounded oxygen anions ($2\times4 = 8$). Upon atomic relaxation the four encased oxygen anions collectively donate four electrons to the four neighboring $Cu^{2+}$ cations converting each of them to a non-magnetic ($d^{10}$) $Cu^{1+}$ cation. This notion can be expressed in formal notation: 2 holes + $(O_4)^{8-}$ + $4Cu^{2+}$ → $(O_4)^{2-}$ + $4Cu^{1+}$. The two remaining out-of-plane electrons from oxygen anions occupy two mutually perpendicular dumbbell-shaped orbitals.

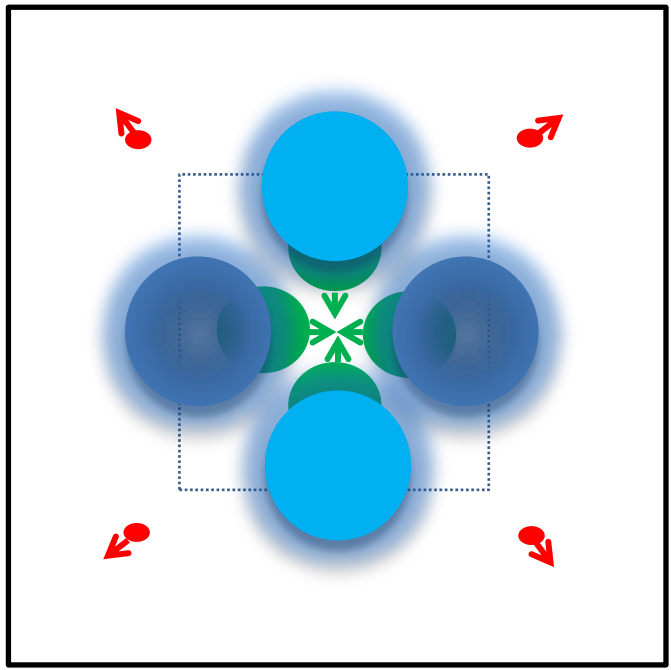

Figure 7. A model for the core of a hole-type large-bipolaron formed in a $CuO_2$ layer is hypothesized. Removing two electrons from the molecular orbitals of four out-of-plane oxygen dianions (large green dots) shifts their equilibrium positions inward and those of the neighboring copper cations (red dots) outward. Collaterally the oxygen anions release four electrons to the four copper cations rendering each of them non-magnetic $Cu^{1+}$ cations. The different shades of blue indicate that the two remaining electrons from the four out-of-plane oxygen orbitals occupy mutually perpendicular dumbbell-shaped molecular orbitals.

At its simplest, the Seebeck coefficient $\alpha$ is the difference of a material's entropy upon adding a bipolaron hole divided by its charge, $2e$.[8] If the charge transfer accompanying large-bipolaron formation eliminates spins on some (e.g. four) spin-½ Cu ions, the Seebeck coefficient garners a contribution proportional to the eliminated magnetic entropy.[3] As schematically illustrated in Fig. 8, the resulting Seebeck coefficient *falls* as the temperature is raised above that of the superconducting transition $T_c$. Were the temperature to be raised high enough for the carrier-free magnetic system to be approximated as paramagnetic, the Seebeck coefficient for large-bipolaron holes would approach:

$$\alpha \rightarrow \frac{k}{2e}\left[ln\left(\frac{1-c}{c}\right) - 4ln(2)\right]. \quad (7)$$

The first contribution within the square brackets is simply the high-temperature limit of the change of the entropy-of-mixing upon adding a large-bipolaron to a system of $n$ large-bipolaron holes distributed among $N$ potential sites with $c \equiv n/N$.[8,9] By contrast, the Seebeck coefficient of a conventional *p*-type superconductor *rises* with increasing temperature once it is raised above $T_c$.

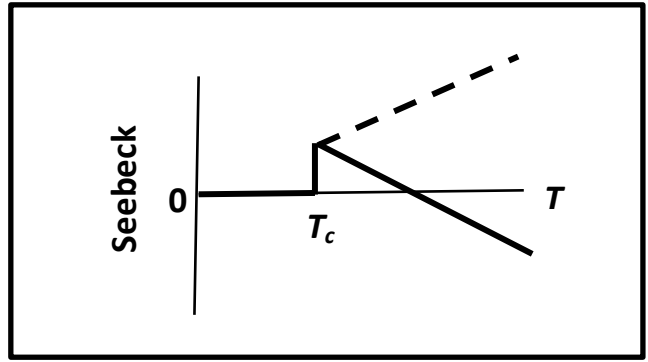

Figure 8. With increasing temperature the normal-state Seebeck coefficient (the entropy transport associated with an added carrier divided by its charge) is progressively driven negative by a large bipolaron's removal of some local magnetic moments and the accompanying magnetic entropy. By contrast, the normal-state Seebeck coefficient for a conventional *p*-type superconductor (dashed line) rises with increasing temperature.

## 4. CONDENSATION TO A LARGE-BIPOLARON LIQUID

An attractive interaction between large bipolarons is required to drive their condensation into a liquid. As in BCS superconductivity, this attractive interaction results from electronic carriers' movements in response to atoms' vibrations.[10,11] BCS envisions these electronic carriers as free. Here, however, these carriers are the self-trapped electronic constituents of large-bipolarons.

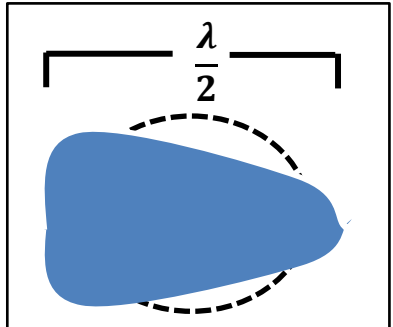

Figure 9. The sloshing of self-trapped carriers in response to atoms' vibrations lowers their effective stiffness constants. This effect is most pronounced for phonons with half-wavelengths $\lambda/2$ that are comparable to the carriers' spatial extent. This local softening of atomic vibrations may even introduce local vibration modes.

Beyond the adiabatic *limit*, in which atoms are frozen at their equilibrium positions, the relaxation of self-trapped carriers in response to atoms' vibrations lowers their stiffness constants.[3] This dynamic effect depicted in Fig. 9 can even introduce local vibration modes (e.g. ghost modes) which principally appear at wavelengths that are less than the self-trapped charges' spatial extent.[12]

Moreover, as illustrated in Fig. 10, self-trapped carriers respond coherently when the phonon wavelengths exceed the separations between large-bipolarons.[13,14] This coherent effect generates an intermediate-range phonon-mediated attraction between large bipolarons which varies roughly as $-\hbar\omega(R_{st}/s)^4$, where $\hbar\omega$ denotes the appropriate phonon zero-point energy while $R_{st}$ and $s$ respectively designate large-bipolarons' characteristic radius and separation.

Short-range repulsions between bipolarons prevent their merger into grander polarons (e.g. quad-polarons).[13,14] These become energetically unfavorable because the Pauli principle requires promoting their electronic carriers above their self-trapping potential well's lowest energy level. Large bipolarons also experience mutual long-range Coulomb repulsions, $(2e)^2/\varepsilon_0 s$.[13,14] Nonetheless, as illustrated in Fig. 11, a net attraction results when the static dielectric constant is sufficiently large. This mutual attraction between large-bipolarons drives their condensing into a liquid.

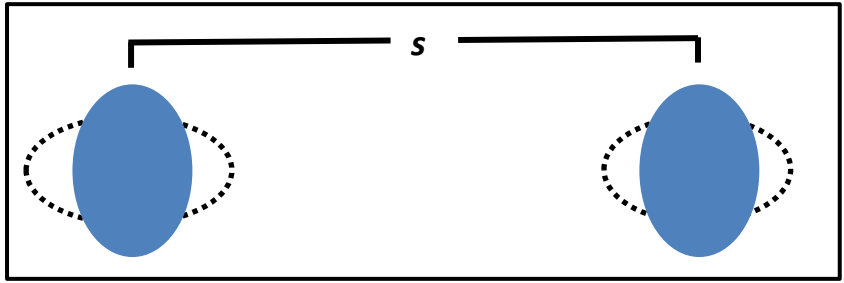

Figure 10. Self-trapped carriers of two large bipolarons separated by distance *s* respond coherently to a longer wavelength phonon. Both self-trapped carriers' shapes are altered from those depicted with dashes to those shown in blue. The net reduction of zero-point phonon energy generated by these coherent responses increases as the separation between large bipolarons is reduced. This effect produces a phonon-mediated attraction.

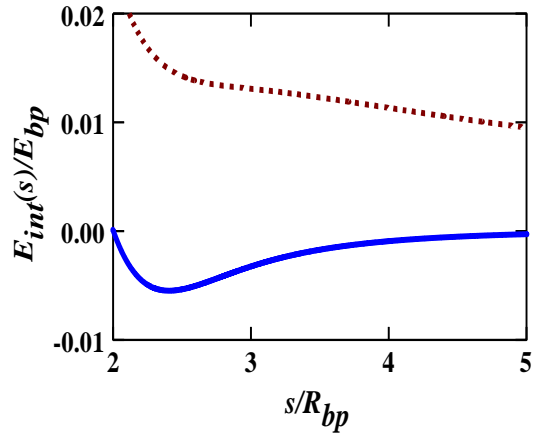

Figure 11. The net interaction energy between large bipolarons $E_{int}(s)$ is plotted as a function of their separation *s*. The dashed red curve shows the intermediate-range phonon-mediated attraction between large bipolarons being overwhelmed by the sum of their short-range and long-range repulsions. The blue curve shows that a shallow attractive minimum develops when a sufficiently large static dielectric constant suppresses the long-range Coulomb repulsions.

Figure 12 shows the phase diagram for large-bipolarons condensing into a bona-fide liquid.[13,14] The condensation temperature increases with rising large-bipolaron density. However, the density of large bipolarons that can form and then condense is also limited as they increasingly compete to displace the ions that surround them.

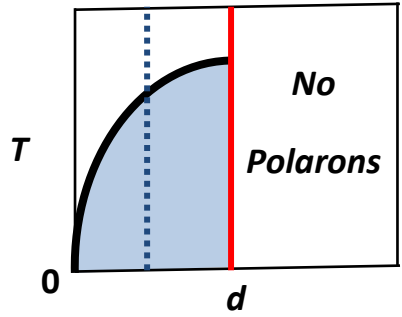

Figure 12. The temperatures $T$ and large-bipolaron densities $d$ within which large-bipolarons condense into a liquid are indicated by the light-blue region of this phase diagram. The red line schematically signifies a density above which neither polarons nor bipolarons can form since they compete to displace the same ions. The dashed dark-blue line schematically indicates a density below which polarons and bipolarons remain bound to the dopants which introduce them.

## 5. A LARGE-BIPOLARON LIQUID'S DISTINCTIVE TRANSPORT AND OPTICAL PROPERTIES

A large bipolaron, a composite quasi-particle comprising a pair of extended self-trapped electronic carriers and their associated pattern of atomic displacements, only moves when these atoms move. As a result, a large-bipolaron's effective bandwidth is very narrow $\sim \hbar\omega(\hbar\omega/E_{bp}) \ll \hbar\omega$ and its velocity is much slower than the characteristic phonon velocity. Indeed, a large-bipolaron's effective mass, $\sim E_{bp}/\omega^2 R_{bp}^2$, is huge compared with the free-electron mass.[3,6]

Excitations of a large-bipolaron liquid are also distinctive. Consider, for example, the excitations for a square-layered large-bipolaron liquid $E(k)$.[15,16] In the long-wavelength limit these excitations are controlled by large-bipolarons' mutual Coulomb repulsions. Then the resulting "plasma-energy" is much less than the phonon energy: $E(0) = \hbar\omega [16\pi(n_2 a^3/b)(\varepsilon_\infty/\varepsilon_0)]^{1/2}$, where $n_2$ denotes the planar density of large bipolarons while $a$ and $b$ respectively represent a square-layer's lattice constant and the interlayer separation and $\varepsilon_\infty/\varepsilon_0 \ll 1$ (e.g. < 0.03 in cuprate insulators). Furthermore, the width of large-bipolarons' excitation spectrum is also much less than the characteristic phonon energy. Concomitantly, the group velocity of these excitations $v(k)$ is also much less than the characteristic phonon velocity. The effective mass associated with these excitations $m_{bpl}$ is also very much larger than the free-electron mass.

These distinctive properties of a large-bipolaron-liquid's excitation spectrum generate unusual electronic transport properties. As examples, the dc resistivity and the frequency-dependent conductivity arising from a large-bipolaron liquid's interaction with ambient acoustic phonons are described.

### 5.1 The dc resistivity

The charge-carrier mobility attributable to a large-bipolaron liquid is $\mu = (2e)\tau/m_{bpl}$. Here $\tau$ denotes the time characterizing momentum transfer between acoustic phonons and the large-bipolaron liquid.[16]

Distinctively, the process via which momentum is transferred between acoustic phonons and excitations of the planar large-bipolaron liquid is unlike that for conventional solids.[16,17] Ordinarily momentum-transfer is associated with processes in which an individual acoustic phonon of wavevector $q$ is absorbed or emitted. Energy conservation for this process requires that $E(k \pm q) - E(k) = \pm\hbar c_s|q|$, where $c_s$ denotes the acoustic-phonon's sound velocity. Upon approximating the difference of the excitation energies by $v(k) \cdot \hbar q$, it becomes clear that conservation of energy cannot be fulfilled for a large-bipolaron liquid since it is precluded by $v(k) \ll c_s$. Thus customary processes do not govern acoustic phonons' scattering of a large-bipolaron liquid.

Figure 13 contrasts the acoustic-phonon scattering of a large-bipolaron liquid with that of conventional charge carriers. In particular, acoustic-phonon scattering of a large-bipolaron liquid is dominated by a two-phonon process.[3,16,17] In this process acoustic phonons are essentially "reflected" from the heavy slow-moving excitations of a large-bipolaron liquid. Impinging phonons cannot penetrate regions of softened vibrations associated with the large-bipolarons. Furthermore, the rate with which acoustic phonons impart momentum to the large-bipolaron liquid is inversely proportional to its effective mass. Stated simply, massive quasi-particles are more difficult to scatter. Thus, the scattering time is proportional to the large-bipolaron-liquid's effective mass: $\tau \propto m_{bpl}$. Therefore, the mobility is *independent* of $m_{bpl}$.[3,16,17]

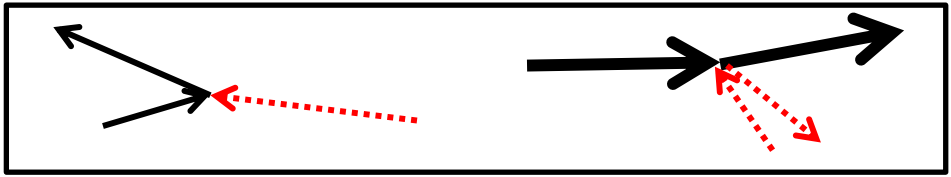

Figure 13. The left side depicts the strong scattering of a light conventional electronic carrier produced by the absorption of an acoustic phonon (dashed red arrow). The illustration on the right side depicts the very weak scattering of a heavy large-bipolaron excitation generated by its "reflection" of a long-wavelength acoustic phonon.

The scattering of ambient acoustic phonons off the nearly dispersion-less excitations of a large-bipolaron liquid is maximized with the equality of their wavevectors. Distinctively, unlike conventional conductors, the scattering rate remains simply proportional to temperature down to a small fraction of the Debye temperature $T_D$: less than $T_D(a/6R_{bp}) \ll T_D$, where $a$ denotes the lattice constant.[16] In particular, this scattering rate becomes[16]

$$\frac{1}{\tau} \sim \omega_D \left(\frac{\kappa T}{E_{bp}}\right)\left(\frac{a}{R_{bp}}\right), \qquad (8)$$

where $\omega_D$ denotes the acoustic-phonon's Debye frequency. Thus, the scattering time $\tau$ is seen to be less than the characteristic vibration period since $\kappa T/E_{bp} \ll 1$ and $a/R_{bp} \ll 1$. The concomitant in-plane mobility and resistivity in the layered-large-bipolaron liquid expressed in terms of the sound velocity $c_s$ and the density of large bipolarons $n_{bp}$ are then[16]

$$\mu = \frac{(2e)\tau}{m_{lbl}} \sim \frac{(2e)c_s R_{bp}}{\kappa T} \qquad (9)$$

and

$$\rho \sim \frac{1}{n_{bp}(2e)^2} \frac{\kappa T}{c_s R_{bp}}. \qquad (10)$$

With reasonable estimates of the relevant physical parameters the mobility and dc resistivity are comparable to experimental values (e.g. ~1 cm²/V-sec and ~400 μΩ-cm respectively at 300 K in $Sr_{0.2}La_{0.8}CuO_4$).

**5.2 The frequency-dependent conductivity**

The scattering time $\tau$ for a large-bipolaron liquid is orders of magnitude longer than that for a conventional conductor.[16] As shown in Eq. (8), the scattering time for a large-bipolaron liquid is even longer than the reciprocal of the Debye frequency. Thus, Figure 14 shows the Drude-like fall-off of the large-bipolaron- liquid's conductivity with applied frequency $\Omega$ occurring at a very much lower frequency, $\Omega \sim 1/\tau$, than for a conventional conductor: $\sigma(\Omega) = \sigma(0)/[1 + (\Omega\tau)^2]$. Indeed, the Drude-like fall off of a large bipolaron occurs below the frequency of the phonons which scatter it.[16]

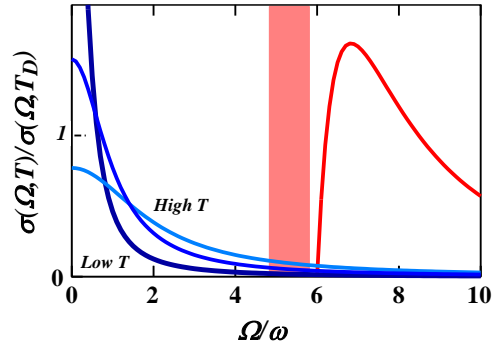

Figure 14. The frequency-dependent conductivity of a large-bipolaron liquid $\sigma(\Omega,T)$ has two distinct components. Large-bipolarons' collective motion generates the temperature-dependent Drude-like contributions (depicted by blue cuves) at applied frequencies $\Omega$ that are primarily below the characteristic phonon frequency $\omega$. In addition, optical liberation of large-bipolarons' self-trapped electronic carriers produces the asymmetric band (plotted in red) at frequencies well above $\omega$. In addition, optical excitation of self-trapped carriers to excited states within their self-trapping potential wells can produce several relatively narrow bands at somewhat lower frequencies (contained within the figure's red band). The "gap" between the low- and high-frequency contributions opens as the temperature is lowered.

Distinctively, a second contribution to large-bipolarons' frequency-dependent conductivity occurs at higher frequencies than those of the characteristic phonons. This contribution arises from excitations of large-bipolarons' self-trapped electronic carriers.[18] The liberation of electronic carriers from the potential well within which they are self-trapped generates a broad asymmetric absorption band. The asymmetry of a large-polaronic absorption is different from that of a small-polaronic absorption. A large-polaron absorption falls off more slowly as the frequency is raised above that of its peak.[16,18] By contrast, a small-polaron absorption falls off more slowly as the frequency is lowered below that of its peak.[19] In addition, excitations of a large-bipolaron's self-trapped electronic carriers to higher-lying states within their self-trapping potential well can produce additional relatively narrow absorptions below the broad absorption band.[3,16,18]

Furthermore, since $\tau$ is inversely proportional to the temperature $T$, the drop-off of the Drude conductivity shifts to ever lower frequencies as the temperature is lowered. As a result, lowering the temperature opens a "gap" between large-bipolarons' high- and low-frequency absorptions.[16]

## 6. BIPOLARONIC SUPERCONDUCTIVITY

Super-flow, superfluidity or superconductivity, occurs when mutually interacting mobile particles undergo a Bose condensation into a collective ground-state which supports non-dissipative irrotational flow. The Bose condensation ensures finite occupation of this ground-state. The dominance of the system's properties by its ground-state ensures that it carries zero entropy. As such, a superconducting state's Seebeck coefficient vanishes since it is proportional to the transported entropy. Non-dissipative flow results when excitations that would impart resistance to the collective ground-state's flow are suppressed. "Rigidity" of interacting particles' ground-state enforces its irrotational flow. This irrotational flow precludes carriers' orbital paramagnetism thereby generating superconductors' Meissner effect.

Bipolaronic superconductivity envisions electronic carriers self-trapping to form mobile individual singlet pairs. That is, bipolarons' electronic carriers are non-overlapping real-space singlet pairs. As such, bipolarons are analogous to the individual atoms of the superfluid phase of liquid $^4$He. Indeed, bipolarons move very slowly with extremely large effective masses since their motion requires significant movement of the surrounding atoms. All told, bipolaronic superconductivity is taken as analogous to the superfluidity of $^4$He albeit with charged particles.

As with polarons, there are two distinct types of bipolarons. The self-trapped electronic carriers of a small bipolaron collapse to a single site. By contrast, a large bipolaron's self-trapped electronic carriers extend over multiple sites.

Self-trapped carriers in multi-dimensional electronic systems always collapse to a single site when governed by their short-range interactions with surrounding atoms, akin to covalent semiconductors' deformation potential. Thus, short-range electron-phonon interactions foster small-bipolaron formation. By contrast, self-trapped carriers' Coulomb-interaction-based long-range interactions with surrounding ions generally foster large-bipolaron formation.

Small-bipolaron motion, like that of small polarons, is generally incoherent. Coherence is destroyed when the energy change in moving between sites exceeds the associated transfer energy. Thus, small polarons and bipolarons are generally observed to move with very low mobility (<< 1 cm$^2$/V-sec) via a succession of thermally assisted hops. Small-bipolarons' pairs even break apart and jump separately when the temperature is high enough for atomic vibrations to be treated as classical.[20] Although an ideal crystal will support coherent motion in the low-temperature limit, even minimal disorder obliterates coherence rendering small polarons and bipolarons immobile. Immobile carriers are not a suitable basis for the collective resistance-less flow of a superconducting condensate.

For example, small-bipolaron formation and motion has been studied in boron carbides over an extremely wide temperature range, 0.1 K to ~ 2000 K.[21,22] These materials, B$_{12+x}$C$_{3-x}$ for $0.1 < x < 1.7$ contain $x/2$ paired holes that execute low-mobility thermally assisted hopping between twelve-atom boron-rich icosahedra. A search for evidence of superconductivity found none down to 0.1 K.[23]

Large bipolarons form when electronic charge carriers interact strongly with especially displaceable ions. Large bipolarons, like large polarons, move coherently with moderate mobilities (~ 1 cm$^2$/V-sec at 300 K) that fall with increasing temperature as their very weak scattering by phonons compensates for their very large effective masses. Distinctively, the multi-atomic extent of a large bipolarons' electronic carriers renders their motion relatively impervious to micro-structural defects.

With a large enough static dielectric constant, the phonon-mediated attraction between large bipolarons can drive their condensation into a liquid. Just as superfluid $^4$He is the fluid Bose condensate of its liquid, so large-bipolaron superconductivity is the fluid Bose condensate of its liquid.

Although the excitations of liquid $^4$He are its sound waves because its atoms are charge-neutral, the electronic charges of large bipolarons produce excitations similar to plasma oscillations.[1-3,16] Distinctively, the huge effective mass of a large bipolaron $m_b$ and the exceptionally large static dielectric constant $\varepsilon_0$ which underlies its formation severely limits the energy and dispersion of its liquid's excitations. In particular, both the plasma energy $\hbar\omega_p \equiv [4\pi n_{bp}(2e)^2/\varepsilon_0 m_{bpl}]^{1/2}$ and the dispersion of the large-bipolaron excitation spectrum are generally less than the characteristic atomic vibration energy.[16] Although the excitation spectra of liquid $^4$He and of a large-bipolaron liquid differ from one another, they both satisfy the Landau condition for the friction-less flow of their ground-states.[1-3]

Super-flow results from the finite occupation of fluid ground-states of these liquids' Bose condensates. The Bose condensation temperature of a large-bipolaron liquid $T_c$ is of the order of the temperature corresponding to its plasma energy, $T_c \sim \hbar\omega_p/\kappa \propto (n_{bp}/\varepsilon_0 m_{bpl})^{1/2}$. This feature is consistent with several observations of novel oxide superconductors.[24] Since $m_{bpl}$ is inversely proportional to the volume of a large bipolaron, $T_c$ increases as a CuO$_2$-based large-bipolaron encompasses progressively more contiguous CuO$_2$ layers of a layered cuprate superconductor. $T_c$ decreases with the addition of insulating layers between contiguous CuO$_2$ layers of a cuprate superconductor since the large-bipolaron density $n_{bp}$ then falls. The extremely large value of $\varepsilon_0$ (20,000 at 1 K) keeps the superconducting transition temperature of doped $SrTiO_3$ below 1 K even when its carrier density is increased to $10^{21}$ cm$^{-3}$.

The large-bipolaron liquid is analogous to liquid $^4$He. Superfluidity, the frictionless flow of liquid $^4$He, is generally attributed to the Bose condensation of liquid $^4$He into a ground-state that remains liquid rather than solidifying. As depicted in Fig. 15, the loss of superconductivity in doped $La_2CuO_4$ with 1/8 hole per $CuO_2$ structural unit suggests global solidification of bipolarons. In particular, this doping corresponds to bipolarons ordering in a manner commensurate with the $CuO_2$-layer's square planar lattice: $1/8 = 2/(4\times4)$. Away from carrier concentrations which permit global solidification commensurate with the underlying lattice, non-solidified remnants of the large-bipolaron liquid's Bose condensate are presumed to produce its superconductivity.

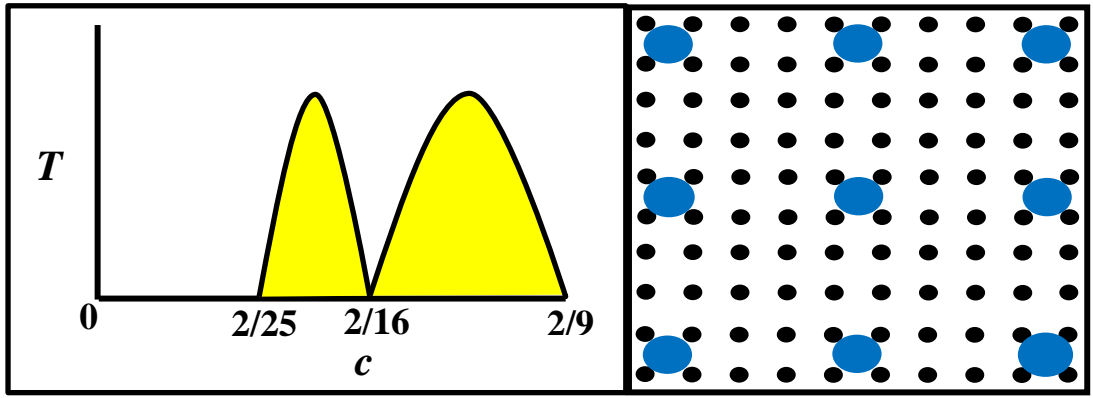

Figure 15. A schematic representation of the superconducting regions of doped $La_2CuO_2$ (yellow) is shown on the left, where the transition temperature $T_c$ is plotted versus the nominal concentration of holes $c$. The cores of the ordered arrangement of large bipolarons (blue) corresponding to $c = 1/8$ is illustrated on the right.

## 7. ADDITIONAL DISTINGUISHING PHENOMENA

This large-bipolaron approach primarily requires modest carrier densities being introduced into materials with significant densities of exceptionally displaceable ions. Empirically, these materials are characterized by very large static dielectric constants, $\varepsilon_0/\varepsilon_\infty \gg 2$, and modest high-frequency dielectric constants, $\varepsilon_\infty \sim 2 - 3$. These features, common to oxide superconductors, are unlike those of conventional superconductors.

Many observations in oxide superconductors suggest that their carrier densities are comparable to those of heavily doped semiconductors rather than to those of conventional conductors. In particular, their superconducting transition temperatures are modulated when the carrier density is modestly altered upon (1) irradiation and (2) carrier injection via the electric-field effect commonly employed in field-effect transistors.

Techniques that take snapshots of materials' electrons see large-bipolarons' self-trapped electronic carriers as localized. Thus large-bipolarons constitute localized centers that foster impinging positrons' annihilation. Concomitantly, large bipolarons' homogenization upon entering the superconducting condensate progressively decreases positrons' annihilation. By contrast, positron annihilation rates are unaltered at a conventional superconductor's transition.

The defining feature of large-bipolarons' self-trapped electronic carriers is that they establish the localized states within which they are trapped. Thus, the chemical potential associated with self-trapped electronic carriers remains pinned between their energy and that of vacant states. By contrast, the chemical potential of a non-polaronic material shifts with the carrier density.

Since large-bipolaron formation requires significant displacements of surrounding ions, large-bipolarons cannot exist in conventional metals. Thus, large-bipolarons' superconducting condensate should not be able to penetrate into a metal to produce a conventional proximity effect. By contrast, large-bipolarons' superconducting condensate may be able to penetrate into or transit through insulating materials for which $\varepsilon_0 \gg 2\varepsilon_\infty$ to produce distinctive proximity and tunneling effects.

## 8. SUMMARY

Exceptionally large ratios of a semiconductor's static to optical dielectric constants enable its electronic charge carriers' pairing to form large singlet bipolarons. Furthermore, Coulomb interactions between a planar large-bipolarons' two electronic carriers induces it to assume a four-lobed shape. It is hypothesized that the charge redistribution associated with bipolarons' displacements of surrounding magnetic ions could eliminate their spins. This circumstance produces an unusual Seebeck coefficient.

The dynamic responses of large-bipolarons' self-trapped electronic carriers to atoms' vibrations reduce their frequencies. Thus large-bipolarons can introduce local vibration modes, "ghost modes." Distinctively, these vibration modes are limited to phonon half-wavelengths that exceed a large-bipolaron's diameter. Moreover, the collective dynamic responses of multiple large-bipolarons to longer wavelength phonons produce a phonon-mediated attractive interaction. This modest attractive interaction can drive large-bipolarons' condensation into a liquid if the static dielectric constant is large enough to severely limit large-bipolarons' mutual Coulomb repulsions.

Distinctive transport and optical properties of a large-bipolaron liquid facilitate its identification. The very weak acoustic-phonon scattering of slow moving heavy massed large-bipolarons results in their possessing a moderate mobility that varies inversely with temperature even well below the Debye temperature. The extremely weak scattering of large bipolarons also confines their temperature-dependent Drude-like responses primarily to frequencies below the Debye frequency. In addition, the optically induced excitation of large bipolarons' self-trapped electronic carriers produces broad asymmetric absorption bands at frequencies well above the Debye frequency. The Bose condensation of a large-bipolaron liquid can yield superconductivity if its ground-state remains fluid rather than solidifying.